\documentstyle[12pt]{article}
\begin{document}

\def\tn{\otimes} \def\nc{\newcommand} \def\ep{\varepsilon} \def\t{\tau}
\def\ph{\varphi} \def\ps{\psi} \def\inv{^{-1}} \def\ov{\over} \def\la{\lambda}
\def\iy{\infty} \def\ch{\raisebox{.3ex}{$\chi$}} 
\renewcommand{\sp}{\vspace{1ex}} \def\be{\begin{equation}} \def\ee{\end{equation}}
\def\tr{{\rm Tr}\;} \def\hf{{1\ov2}} 
\nc{\twotwo}[4]{\left(\begin{array}{cc}#1&#2\\&\\#3&#4\end{array}\right)}
\def\iy{\infty} \def\r{\rho}  \def\A{{\rm Airy}}
\def\pl{\partial} \def\tn{\otimes}

\hfill May 11, 2004
\begin{center}{\bf Matrix Kernels for the Gaussian Orthogonal and Symplectic Ensembles}\end{center}
\begin{center}{{\bf Craig A.~Tracy}\\
{\it Department of Mathematics \\
University of California\\
Davis, CA 95616, USA}}\end{center}
\begin{center}{{\bf Harold Widom}\\
{\it Department of Mathematics\\
University of California\\
Santa Cruz, CA 95064, USA}}\end{center}
\sp
\begin{abstract} We derive the limiting matrix kernels for the the Gaussian Orthogonal and Symplectic 
ensembles scaled at the edge, with proofs of convergence in the operator norms
that assure convegence of the determinants.
\end{abstract}

\renewcommand{\theequation}{1.\arabic{equation}}
\begin{center}{\bf I. Introduction}\end{center}

For a large class of finite $N$ determinantal processes  the limiting
distribution, as $N\to\iy$, 
of the right-most ``particle'' is expressible as the Fredholm determinant 
of an operator $K_{\A}$ with kernel
\[{A(x)A'(y)-A'(x)A(y)\ov  x-y}=\int_0^\iy A(x+z)\,A(y+z)\,dz,\]
where $A(x)={\rm Ai}(x)$ is the Airy function.
(See \cite{Joh,TW3} for recent reviews.) Typically these results are proved
by first establishing that for finite $N$ the distribution of the the right-most particle
is the Fredholm determinant of an operator $K_N$.  The limit theorem then follows once one
proves $K_N\to K_{\A}$ in trace norm.  The classic example is the finite
$N$ Gaussian Unitary Ensemble (GUE) where $K_N$ is the Hermite kernel \cite{mehta}
and the limit $N\to\iy$ is the edge scaling limit of the largest eigenvalue.
In this case the unitary invariance of the underlying probability measure is manifest.

As is well known one can also consider the edge scaling of the largest eigenvalue for
the Gaussian Orthogonal Ensemble (GOE) or the Gaussian Symplectic Ensemble (GSE) \cite{TW1}.
These limiting distributions are now widely believed to describe the universal behavior of
a right-most particle for a large class of processes under additional symmetry restrictions 
\cite{baikRains,PS}. In the orthogonal
and symplectic ensembles, the distribution of the largest eigenvalue is the square root
of a Fredholm determinant of an operator $K_N$ 
whose kernel is now a $2\times 2$ matrix kernel (of course, different $K_N$ for different ensembles).
In \cite{TW1} the derivation of the 
limiting distributions for the largest eigenvalue did not use limiting kernels, as it
did for the ensembles scaled in the bulk. Instead, differential equations determining the the 
finite $N$ distributions were obtained first and then a limiting argument was applied to these.

The edge-scaled kernels are already in the literature \cite{FNH,F}, but
as far as we know limit theorems in the operator norms that give convergence of the determinants are 
not. Because of the 
current interest in these ensembles, it seems useful to state and prove such  
theorems for the matrix kernels. For GSE we shall establish convergence 
in trace norm to a limiting operator $K_{\rm GSE}$. The GOE case is more awkward and for
that we shall use weighted Hilbert spaces and an extension of the determinant involving the 
regularized 2-determinant, and convergence 
to $K_{\rm GOE}$ will be in a combination of Hilbert-Schmidt and trace 
norms. These will give convergence of the determinants.

\vspace{2ex}
\renewcommand{\theequation}{2.\arabic{equation}}
\begin{center}{\bf II. The finite \boldmath $N$ kernels}\end{center}

We shall follow the notation of \cite{TW2}, more or less. For both finite $N$ ensembles
we denote by $K_N(x,\,y)$ the matrix kernel (and $K_N$ the corresponding operator) such
that the expected value of $\prod_k(1+f(\la_k))$ is given by
\be E\left(\prod_k(1+f(\la_k))\right)=\sqrt{\det\,(I+K_N\,f)}.\label{EV}\ee
Here $f$ is a ``general'' function and in the determinant it denotes multiplication by $f$.
In particular the probability of finding no eigenvalues in a set $J$ is given by
\be \sqrt{\det\,(I-K_N\,\ch_J)}.\label{Prob}\ee

Here are the formulas for $K_N$. We set $\ep(x)=\hf\,{\rm sgn}\,x$, denote by $\ep$ the
operator with kernel $\ep(x-y)$, and denote by $\ph_n$ the harmonic oscillator wave 
functions.

For GSE, with weight function $e^{-x^2}$ and $N$ odd, we define
\[S_N(x,y)=\sum_{n=0}^{N-1}\,\ph_{n}(x)\,\ph_{n}(y)+\sqrt{{N\ov2}}\;\ph_{N-1}(x)\;
\ep\ph_{N}(y),\]
\[IS_N(x,y)=\sum_{n=0}^{N-1}\,\ep\ph_{n}(x)\;
\ph_{n}(y)+\sqrt{{N\ov2}}\;\ep\ph_{N-1}(x)\;\ep\ph_{N}(y),\]
\[S_ND(x,y)=-\sum_{n=0}^{N-1}\,\ph_{n}(x)\,\ph_{n}'(y)-\sqrt{{N\ov2}}\;\ph_{N-1}(x)\,
\ph_{N}(y).\]
Then
\be K_N(x,y)=\hf\,\twotwo{S_N(x,y)}{S_ND(x,y)}{IS_N(x,y)}{S_N(y,x)}.\label{KNGSE}\ee

For GOE, with weight function $e^{-x^2/2}$ and $N$ even we define
$S_N(x,y),\ IS_N(x,y)$ and $S_ND(x,y)$ by exactly the same formulas and then
\be K_N(x,y)=\twotwo{S_N(x,y)}{S_ND(x,y)}{IS_N(x,y)-\ep(x-y)}{S_N(y,x)}.\label{KNGOE}\ee

Despite their apparent similarity the kernels for GSE and GOE have very different 
properties. First, the functions $\ep\ph_{n}$, which do not vanish at $\pm\iy$ when 
$n$ is even, make it appear that the entries of $K_N(x,\,y)$ do not vanish at $\pm\iy$, but
they are actually exponentially small for GSE. (See derivation in Section 8 of \cite{TW2}.) Thus
the $f$ in (\ref{EV}) can be quite general and even grow at $\pm\iy$, and 
$K_Nf$ is a trace class operator on $L^2\oplus L^2$. 

This is not the case for GOE. First, the entries do not all vanish at $\pm\iy$; second, the discontinuity in 
$\ep(x-y)$ prevents $K_Nf$ from being a trace class operator. (The discussion in Section 9 
of \cite{TW2} skirted these issues.) To take care of these problems we use weighted
$L^2$ spaces and a generalization of the determinant. Let $\r$ be any weight function such that
$\r\inv\in L^1$ and such that all $\ph_n\in L^2(\r)$.\footnote{Recall that $L^2(\r)
=\{h:\int |h(x)|^2\r(x)dx<\iy\}$.} For any such $\r$ the matrix $K_N$ is a Hilbert-Schmidt
operator on $L^2(\r)\oplus L^2(\r\inv)$,\footnote{That $\ep$ is a Hilbert-Schmidt operator from 
$L^2(\r)$ to $L^2(\r\inv)$ is equivalent to $\r\inv\in L^1$.} and this is
the space on which $K_N$ acts. The diagonal entries of $K_N$ are
finite rank, hence trace class. Now the definition of determinant extends to Hilbert-Schmidt 
operator matrices $T$ with trace class diagonal entries by setting
$\det(I-T)=\det_2(I-T)\,e^{-{\rm tr}\,T}$, where ${\rm tr}\,T$ denotes the sum of the traces of the 
diagonal entries of $T$ and $\det_2$ is the 
regularized 2-determinant.\footnote{
If $T$ is a Hilbert-Schmidt operator with eigenvalues $\mu_k$ then $\det_2\,(I-T)
=\prod (1-\mu_k)\,e^{\mu_k}$. 
See \cite{GK}, Sec. IV.2.} It follows from an identity for 2-determinants
(\cite{GK}, p.$\,169$) that for this extended definition we still have the relation
\[\det\,(I-T_1)\,(I-T_2)=\det\,(I-T_1)\,\det\,(I-T_2).\]
Using this fact, and defining $\det\,(I-K_Nf)$ in this way,\footnote{The so-defined $\det\,(I-K_Nf)$ is independent of the choice of $\r$, since
eigenfunctions corresponding to nonzero eigenvalues belong to $L^2(\r)\oplus L^2(\r\inv)$ for all 
permissible $\r$. In fact it is given by the usual Fredholm expansion, modified for
matrix kernels. We prefer the operator definition since it will enable us
to establish convergence of the determinants more easily.}
the discussion of \cite{TW2} can be carried through for any $f\in L^\iy$.

\sp
\renewcommand{\theequation}{3.\arabic{equation}}
\begin{center}{\bf III. Scaling GSE at the edge}\end{center}

Here we follow the notation of Section VII of \cite{TW1} and denote our scaling transformation 
by $\t$, so that
\[\t(x)=\sqrt{2N}+{x\ov\sqrt2 N^{1/6}}.\]
We shall consider only the limit of (\ref{Prob}). We denote here by $J$ any measurable set 
which is bounded below and replace $J$ in (\ref{Prob})
by $\t(J)$. We shall compute the matrix kernel $K_{GSE}$ such that the limit of the determinant
in (\ref{Prob}) as
$N\to\iy$ is equal to $\det\,(I-K_{GSE}\,\ch_J)$.

Observe first that the determinant is unchanged if the kernel $K_N(x,y)\,\ch_{\t(J)}(y)$
is replaced by 
\[\t'\,K_N(\t(x),\,\t(y))\,\ch_J(y),\]
where $\t'=2^{-1/2}N^{-1/6}$. It is also unchanged if the upper-right entry
is multiplied by $\t'$ and the lower-left entry divided by $\t'$. (We shall use the same
notation for the modified kernel.) Thus it is a matter of 
determining the limit, in trace norm, of the entries of the modified matrix kernel.

If we write
\[\ph(x)=\left({N\ov2}\right)^{1/4}\,\ph_N(x),\quad\ps(x)=\left({N\ov2}\right)^{1/4}\,\ph_{N-1}(x),\]
and set
\[S_N^0(x,\,y)=\sum_{n=0}^{N-1}\,\ph_{n}(x)\,\ph_{n}(y),\]
then 
\[S_N(x,\,y)=S_N^0(x,\,y)+\ps(x)\,\ep\ph(y).\]

Formula (57) of \cite{TW1} is in our notation
\[ S_N^0(x,y)=\int_0^\iy [\ph(x+z)\,\ps(y+z)+\ps(x+z)\,\ph(y+z)]\,dz.\]
Define $\ph_\t=\ph\circ\t$ and $\ps_\t=\ps\circ\t$. Then the substitution $z\to\t' z$ in 
the integral shows that
\[\t'\,S_N^0(\t(x),\,\t(y))={\t'}^2\int_0^\iy [\ph_\t(x+z)\,\ps_\t(y+z)+\ps_\t(x+z)\,\ph_\t(y+z)]\,dz.\]
It follows from results on the asymptotics of Hermite polynomials that 
\be\lim_{N\to\iy}\t'\,\ph_{\t}(x)=\lim_{N\to\iy}\t'\,\ps_{\t}(x)={1\ov\sqrt2}A(x)\label{lims}\ee
pointwise, where $A(x)$ denotes the Airy function Ai($x$), and that there are estimates
\be \t'\,\ph_{\t}(x)=O(e^{-x}),\quad \t'\,\ps_{\t}(x)=O(e^{-x})\label{ests}\ee
which hold uniformly in $N$ and for $x$ bounded below. (There is a better bound but this one 
is good enough for our purposes. See \cite{Ol}, p. 403.) We shall show that this implies that if
\[K_\A(x,\,y)=\int_0^\iy A(x+z)\,A(y+z)\,dz,\]
then
\[\t'\,S_N^0(\t(x),\,\t(y))\to K_\A(x,\,y)\]
in trace norm on any space $L^2(s,\,\iy)$.

To show that
\[{\t'}^2\int_0^\iy \ph_\t(x+z)\,\ps_\t(y+z)\,dz\to \hf K_\A(x,\,y),\]
(the other half of $\t'\,S_N^0(\t(x),\,\t(y))$ is treated similarly) we write the difference as
\[\int_0^\iy [\t'\ph_\t(x+z)-2^{-1/2}A(x+z)]\,\t'\ps_\t(y+z)\,dz
+\int_0^\iy 2^{-1/2}A(x+z)\,[\t'\ps_\t(y+z)-2^{-1/2}A(y+z)]\,dz.\]
Each summand is an integral over $z$ of rank one kernels, and the trace norm of an integral
is at most the integral of the trace norms. Thus the trace norm of the first summand is at most
\be\int_0^\iy \|\t'\ph_\t(x+z)-2^{-1/2}A(x+z)\|_2\;\|\t'\ps_\t(y+z)\|_2\;dz,\label{GSEnormest}\ee
where the first $L^2$ norm is taken with respect to the $x$ variable and the second with respect to
the $y$ variable. It follows from (\ref{ests}) that the second norm is $O(e^{-z})$ uniformly
in $N$, and from
(\ref{ests}) and (\ref{lims}) that the first norm is also $O(e^{-z})$ uniformly in $N$ and tends
pointwise to zero. Thus the integral has limit zero. The same argument applies to the othe
integral.

Thus $S_N^0$ scales to $K_\A$ and it remains to consider $\t'\,\ps_\t(x)\,(\ep\ph)_\t(y)$,
the last summand in $\t'\,S_N(\t(x),\,\t(y))$. Of course $\t'\,\ps_\t(x)\to 
{1\ov\sqrt2}A(x)$ in trace norm. Define
\[c_\ph=\hf\int_{-\iy}^\iy\ph(x)\,dx,\ \ \ c_\ps=\hf\int_{-\iy}^\iy\ps(x)\,dx,\]
\[\Phi_\t(x)=\int_x^\iy \t'\phi_\t(z)\,dz,\ \ \ \Psi_\t(x)=\int_x^\iy \t'\ps_\t(z)\,dz.\]
We have
\[(\ep\ph)_\t(y)=c_\ph-\int_{\t(y)}^\iy \ph(z)\,dz=-\int_y^\iy\t'\ph_\t(z)\,dz=-\Phi(y).\]
(Since $N$ is odd so is $\ph$, so $c_\ph=0$.)
{}From (\ref{lims}) and (\ref{ests}) it follows that this converges to
\[-{1\ov\sqrt2}\int_y^\iy A(z)\,dz\]
in the norm of any space $L^2(s,\,\iy)$. 

Thus we have shown that 
\[\t'\,S_N(\t(x),\,\t(y))\to K_\A(x,\,y)-\hf\,A(x)\,\int_y^\iy A(z)\,dz\]
in the trace norm of operators on any space $L^2(s,\,\iy)$.

Next, observe that $S_ND(x,\,y)=-\pl_y S_N(x,\,y)$ and so (recall that we multiply the 
upper-right corner by $\t'$)
\[{\t'}^2\,S_ND(\t(x),\,\t(y))=-\pl_y \t'S_N(\t(x),\,\t(y)).\]
Since the limiting relation we found above for $S_N$ holds even after taking $\pl_y$ we see that
\[{\t'}^2\,S_ND(\t(x),\,\t(y))\to -\pl_y K_\A(x,\,y)-\hf A(x)\,A(y)\]
in trace norm as $N\to\iy$.

If we recall that we divide the 
lower-left corner by $\t'$ we see that it remains to find the limit of
\[IS_N(\t(x),\,\t(y))=\ep S_N^0(\t(x),\,\t(y))+\ep\ps(\t(x))\,\ep\ph(\t(y)).\]
The last term equals
\[-(c_\ps-\Psi_\t(x))\,\Phi_\t(y).\]
If we make the substitution $z\to \t' z$ in the integral for $\ep S_N^0(\t(x),\,\t(y))$ it becomes
\be\int_0^\iy [(\ep\ph_\t)(x+z)\,\t'\ps_\t(y+z)+(\ep\ps)_\t(x+z)\,\t'\ph_\t(y+z)]\,dz.\label{Iint}\ee
Replacing $\ep\ph_\t$ and $\ep\ps_\t$ by what they are in terms of $\Phi_\t$ and $\Psi_\t$ gives
\[-\int_0^\iy [\Phi_\t(x+z)\,\t'\ps_\t(y+z)+\Psi_\t(x+z)\,\t'\ph_\t(y+z)]\,dz+c_\ps\Phi_\t(y).\]
Putting these together we obtain
\[IS_N(\t(x),\,\t(y))=-\int_0^\iy [\Phi_\t(x+z)\,\t'\ps_\t(y+z)+\Psi_\t(x+z)\,\t'\ph_\t(y+z)]\,dz
+\Psi_\t(x)\,\Phi_\t(y).\]
Now we use (\ref{lims}) and (\ref{ests}) as before, and deduce that
\[IS_N(\t(x),\,\t(y))\to-\int_x^\iy K_A(z,y)\,dz+\hf\int_x^\iy A(z)\,dz\,\cdot\,\int_y^\iy A(z)\,dz\]
in trace norm.

Thus if we set
\[S(x,\,y)=K_\A(x,\,y)-\hf\,A(x)\,\int_y^\iy A(z)\,dz,\]
\[SD(x,\,y)=-\pl_y K_\A(x,\,y)-\hf A(x)\,A(y),\]
\[IS(x,\,y)=-\int_x^\iy K_A(z,y)\,dz+\hf\int_x^\iy A(z)\,dz\,\cdot\,\int_y^\iy A(z)\,dz,\]
then we have shown that for GSE
\[\t'K_N(\t(x),\,\t(y))\to K_{\rm GSE}(x,\,y):=\hf\,\twotwo{S(x,y)}{SD(x,y)}{IS(x,y)}{S(y,x)}\]
in trace norm as $N\to\iy$. 

In particular, the probability that no eigenvalue
lies in $\t(J)$ tends to
\[\sqrt{\det\,(I-K_{\rm GSE}\,\ch_J)}\]
as $N\to\iy$.

\pagebreak
\renewcommand{\theequation}{4.\arabic{equation}}
\begin{center}{\bf IV. Scaling GOE at the edge}\end{center}

Before computing limits let us see what the norm of a rank one kernel $u(x)v(y)$ is when 
thought of as taking a space 
$L^2(\r_1)$ to a space $L^2(\r_2)$. The operator, denoted by $u\tn v$, takes a function $h\in L^2(\r_1)$ to $u\,(v,\,h)$, 
and so its norm is the $L^2(\r_2)$ norm of $u$ times the norm of $v$ in the space dual to $L^2(\r_1)$,
which is  $L^2(\r_1\inv)$. Thus 
\be\|u\tn v\|=\|u\|_{L^2(\r_2)}\,\|v\|_{L^2(\r_1\inv)}.\label{tensornorm}\ee

We determine the limit of the scaled kernel 
\[\t'K_N(\t(x),\t(y))\]
where $K_N$ is now given by (\ref{KNGOE}).
The space on which this acts is $L^2(\r)\oplus L^2(\r \inv)$ with any fixed permissible $\r$ which we now
assume has
at most polynomial growth at $+\iy$.\footnote{All functions are thought of as defined on 
some interval $(s,\,\iy)$.} As before, we
multiply the upper-right corner by $\t'$ and divide the lower-left corner by $\t'$, but do not
change notation.

For $\t'S_N^0(\t(x),\,\t(y))$, the main part of $\t'S_N(\t(x),\,\t(y))$, consider the analogue of
(\ref{GSEnormest}). Since this kernel takes $L^2(\r)$ to itself (\ref{tensornorm}) gives
for the analogue here
\be\int_0^\iy \|\t'\ph_\t(x+z)-2^{-1/2}A(x+z)\|_{L^2(\r)}\;\|\t'\ps_\t(y+z)\|_{L^2(\r\inv)}\;dz.
\label{GOEnormest}\ee
The second factor is $O(e^{-z})$ uniformly in $N$ since $\t'\ps_\t(y+z)=O(e^{-z})$ 
and $\r\inv\in L^1$. The first
factor is at most the square root of
\[\int_s^\iy|\t'\ph_\t(x+z)-2^{-1/2}A(x+z)|^2\r(x)\,dx\]
for some $s$. 
Using (\ref{lims}) and (\ref{ests}) and the fact that $\r$ has at most polynomial growth we see that
this is uniformly $O(e^{-z/2})$ and converges to zero pointwise. Thus (\ref{GOEnormest}) 
has limit zero.

The other integral arising in the aymptotics of $\t'S_N(\t(x),\,\t(y))$ is similar, and we deduce that
\[\t'\,S_N^0(\t(x),\,\t(y))\to K_\A(x,\,y)\]
in trace norm, as an operator on $L^2(\r)$. It remains to consider
$\t'\,\ps_\t(x)\,(\ep\ph)_\t(y)$. First, $\t'\,\ps_\t(x)\to{1\ov\sqrt2}A(x)$ in the space $L^2(\r)$, again
by (\ref{lims}) and (\ref{ests}) and the fact that $\r$ has at most polynomial growth. Now
\[(\ep\ph)_\t(y)=c_\ph-\Phi(y),\]
and since $N$ is even for GOE so is $\ph$ and $c_\ph\neq0$. In fact $c_\ph\to{1\ov\sqrt2}$ 
as $N\to\iy$ and so we see that
\[(\ep\ph)_\t(y)\to{1\ov\sqrt2}-{1\ov\sqrt2}\int_y^\iy A(z)\,dz\]
uniformly, and so in the space $L^2(\r\inv)$. Thus,
\[\t'\,S_N(\t(x),\,\t(y))\to K_\A(x,\,y)+\hf A(x)\left(1-\int_y^\iy A(z)\,dz\right)\]
in trace norm as an operator on $L^2(\r)$. 

Consider now the upper-right corner of the modified matrix,
\[{\t'}^2\,S_ND(\t(x),\,\t(y))=-\pl_y \t'S_N(\t(x),\,\t(y))-{\t'}^2\ps_\t(x)\,\ph_\t(y),\]
whose limit we want to compute as an operator from $L^2(\r\inv)$ to $L^2(\r)$. In the 
application of (\ref{tensornorm}) both norms are that of $L^2(\r)$. In the verification that
\[-\pl_y \t'S_N^0(\t(x),\,\t(y))\to-\pl_y K_\A(x,\,y)\]
we obtain an integral of the form (\ref{GOEnormest}) where the second factor in the integrand
is replaced by
\[\|\t'\ps_\t'(y+z)\|_{L^2(\r)}.\]
This is bounded uniformly in $N$ and $z$ because (\ref{ests}) holds also even after taking derivatives.
The first integrand is estimated as before, so the integral tends to zero in trace norm.

We also have $\t'\ps_\t(x)\to {1\ov\sqrt2}A(x)$ and $\t'\ph_\t(y)\to {1\ov\sqrt2}A(y)$ in $L^2(\r)$, 
and so we find that
\[{\t'}^2\,S_ND(\t(x),\,\t(y))\to -\pl K_\A(x,\,y)-\hf\,A(x)\,A(y)\]
in trace norm as before, but now as operators from $L^2(\r\inv)$ to $L^2(\r)$.

For the lower-left corner, an operator now from $L^2(\r)$ to $L^2(\r\inv)$ so in (\ref{tensornorm})
both norms are $L^2(\r\inv)$ norms, we have as before
\[IS_N(\t(x),\,\t(y))=\ep S_N^0(\t(x),\,\t(y))+\ep\ps(\t(x))\,\ep\ph(\t(y)).\]
The last term equals
\[-\Psi_\t(x)\,(c_\ph-\Phi_\t(y)).\]
Replacing $\ep\ph_\t$ and $\ep\ps_\t$ in (\ref{Iint}) by what they are in terms of $\Phi_\t$ and 
$\Psi_\t$ now gives
\[-\int_0^\iy [\Phi_\t(x+z)\,\t'\ps_\t(y+z)+\Psi_\t(x+z)\,\t'\ph_\t(y+z)]\,dz+c_\ph\Psi_\t(y),\]
so we obtain now the identity
\[IS_N(\t(x),\,\t(y))=-\int_0^\iy [\Phi_\t(x+z)\,\t'\ps_\t(y+z)+\Psi_\t(x+z)\,\t'\ph_\t(y+z)]\,dz\]
\[+c_\ph\,\Psi_\t(y)-\Psi_\t(x)\,(c_\ph-\Phi_\t(y)).\]
Now we deduce that
\[IS_N(\t(x),\,\t(y))\to-\int_x^\iy K_A(z,y)\,dz
+\hf\left(\int_y^x A(z)\,dz+\int_x^\iy A(z)\,dz\,\cdot\,\int_y^\iy A(z)\,dz\right)\]
as trace class operators from $L^2(\r)$ to $L^2(\r\inv)$.

Finally, $\ep(\t(x)-\t(y))=\ep(x-y)$, so this is unchanged. 
 
Thus if we set
\[S(x,\,y)=K_\A(x,\,y)+\hf\,A(x)\left(1-\int_y^\iy A(z)\,dz\right),\]
\[SD(x,\,y)=-\pl_y K_\A(x,\,y)-\hf A(x)\,A(y),\]
\[IS(x,\,y)=-\int_x^\iy K_A(z,y)\,dz+\hf\left(\int_y^x A(z)\,dz+
\int_x^\iy A(z)\,dz\,\cdot\,\int_y^\iy A(z)\,dz\right),\]
then we have shown that for GOE
\[\t'K_N(\t(x),\,\t(y))\to K_{\rm GOE}(x,\,y):=\twotwo{S(x,y)}{SD(x,y)}{IS(x,y)-\ep(x-y)}{S(y,x)}\]
as $N\to\iy$ in the sense that the kernels converge in the Hilbert-Schmidt norm of operators on
$L^2(\r)\oplus L^2(\r\inv)$ and the diagonals converge in trace norm. (In fact
all entries converge in trace norm, except for the fixed Hilbert-Schmidt operator $\ep$.) 

In particular, it follows from the continuity of the 2-determinant in Hilbert-Schmidt norm and 
the trace in trace norm that if the set $J$ is bounded below then the probability that no eigenvalue
lies in $\t(J)$ tends to
\[\sqrt{\det\,(I-K_{\rm GOE}\,\ch_J)}\]
as $N\to\iy$.

\begin{center}{\bf Acknowledgments}\end{center}

This work was supported by National Science Foundation under grants 
DMS-0304414 (first author) and DMS-0243982 (second author).

\end{document}